\documentclass[superscriptaddress,secnumarabic,
amssymb,amsmath,nobibnotes,aps,prd,nofootinbib]{revtex4}%
\usepackage{graphicx}
\usepackage{epsf}
\usepackage{bm}
\usepackage{amsmath}
\usepackage{amsfonts}
\usepackage{amssymb}
\usepackage{epstopdf}
\usepackage{natbib}%
\providecommand{\U}[1]{\protect\rule{.1in}{.1in}}
\newcommand{\be}{\begin{equation}}
\newcommand{\ee}{\end{equation}}

\newcommand{\mincir}{\raise
-3.truept\hbox{\rlap{\hbox{$\sim$}}\raise4.truept\hbox{$<$}\ }}
\newcommand{\magcir}{\raise
-3.truept\hbox{\rlap{\hbox{$\sim$}}\raise4.truept\hbox{$>$}\ }}

\usepackage{color}
\usepackage{float}

\begin{document}
\title{Cosmological solutions in spatially curved universes with adiabatic particle production}

\author{Llibert Arest\'e Sal\'o}
\email{llibert.areste@estudiant.upc.edu}

%\author{Jaume Amor\'os}
%\email{jaume.amoros@upc.edu}
\author{Jaume de Haro}
\email{jaime.haro@upc.edu}
\affiliation{Departament de Matem\`atica Aplicada, Universitat Polit\`ecnica de Catalunya, Diagonal 647, 08028 Barcelona, Spain}

\begin{abstract}
We perform a qualitative and thermodynamic study of two models when one takes into account adiabatic particle production. In the first one, there is a constant particle production rate, which leads to solutions depicting the current cosmic acceleration but without inflation. The other one has solutions that unify the early and late time acceleration. These solutions converge asymptotically to the thermal equilibrium.
\end{abstract}

\maketitle

\section{Introduction}

Several astronomical observations at the end of last century suggested that nowadays the universe is in an accelerated regime
\cite{Riess98, Perlmutter99, Spergel2003, Tegmark2004, Eisenstein2005, Komatsu2011, Planck2015}. There are basically three  theoretical ways to explain this behavior. The first one introduces some fluid or scalar field, generally known as dark energy, with a large negative pressure that violates the strong energy condition (see for a review \cite{CST, AT}). This includes a large number of models, in which the simplest candidate to unify inflation with the current cosmic acceleration is to mix an inflaton field with the cosmological constant \cite{he, haro}.
The second one consists in going beyond Einstein cosmology and considering modified gravity theories such as teleparallelism, $f(R)$ gravity or scalar field theories (see for instance \cite{Clifton:2011jh,Cai:2015emx, odintsov}). One of the main goals of the works that deal with this kind of modified theories is to unify inflation with the current cosmic acceleration \cite{odintsov1}. However, one of their basic defaults is that sometimes the proposed models are very involved and  contain too many parameters. 

\

The last way opts for considering adiabatic particle production, i.e., when the specific entropy (the entropy per particle) is conserved. This effect was treated for first time in 1989 by Prigogine et al. in \cite{Prigogine88}, where the authors successfully insert this macroscopic particle production by adding a negative pressure term into Einstein's field equations. Later, a covariant approach was proposed in \cite{lima92cov}, showing that \cite{Prigogine88} was a particular case of a more general treatment, i.e., when the specific entropy  is considered non constant. Returning to the adiabatic case, despite being the perfect fluid particles produced isentropic, the increase of its number induces an enlargement of the phase space of the system and, hence, a total entropy production \cite{saha}. The main interest of this approach is that there are some models leading to a current cosmic acceleration that succeed in matching with the available observational data \cite{SSL09,LJO10, LB10, JOBL11, LBC12, carneiro2012, LGPB14, CPS14, FPP14, NP15, NP2016, RCN2016,carneiro2016} and even sometimes it is possible to mimic the $\Lambda$CDM model \cite{JOBL11}.
% Now, in principle the rate of particle
%production should be determined from the quantum theory of fields in curved
%spacetime \cite{Birrell}, but since this is not completely developed yet, so
%one need to take account several choices and which are found to be effective
%as described in \cite{SSL09, LJO10, LB10, JOBL11, LBC12, LGPB14, CPS14, FPP14,
%NP15, NP2016, RCN2016}. \newline
Due to its simplicity, adiabatic particle production is generally studied in the flat Friedmann-Lema{\^\i}tre-Robertson-Walker spacetime and only in few works the non flat case is considered \cite{barrow, lima96,lima99, pavon}. In particular, in \cite{HP2016} the authors have studied the dynamics and thermodynamics of a spatially flat universe when the particle production rate is constant. Moreover, in \cite{PHPS2016} the dynamics of a model with a non constant rate, which leads to an early accelerated epoch that mimics inflation and a current cosmic acceleration, has been studied. Even a kinetic formulation for this model has been discussed \cite{lima2014}.

\

For this reason, the aim of the present work is to generalize, at spatially curved cosmologies, the results obtained in those papers. The problem is more involved because in the flat case the dynamics decouples in the sense that it is given by an autonomous first order differential equation of the form $\dot{H}=F(H)$, where $H$ is the Hubble parameter, which in general could be solved analytically. This does not happen when the spacetime is spatially curved because in this case the differential equations for the scale factor, namely $a$, and the Hubble parameter do not decouple, obtaining a two dimensional autonomous dynamical system given by $\dot{H}=F(H,a)$, $\dot{a}=Ha$, which in general is impossible to solve analytically and needs to be studied using the mathematical techniques of dynamical systems, the so-called qualitative analysis.

\vspace{0.5cm}

The work is organized as follows. In section $2$ we deduce for a Friedmann-Lema{\^\i}tre-Robertson-Walker spacetime the equations that model the dynamics of the universe when adiabatic particle production is allowed. Section $3$ is devoted to the dynamical study of the universe when the particle production rate is constant. Here we see that for a closed universe the bounces are not allowed, in contrast with what happens when there is not particle production. In fact, instead of a bounce, the scalar curvature diverges when the Hubble parameter vanishes. The remarkable property is that in the expanding phase all the solutions, at very late time, have an effective Equation of State parameter that converges to $-1$, meaning that all of them depict a universe that at late times enters in a de Sitter regime, which could model the current cosmic acceleration. The case of variable production rate is studied in section $4$, where we basically show that for all the cases there are orbits unifying inflation with the current cosmic acceleration, that is, solutions of the dynamical equations that depict a universe with an early and late time acceleration. Finally, in last section we study the thermodynamical properties of the models studied in the previous sections.

\

We will use natural units: $c=\hbar=8\pi G=k_B=1$.

\

\section{Cosmological models driven by particle production}

Assuming homogeneity and isotropy at large scales, our universe is well described by the so-called Friedmann-Lema\^itre-Robertson-Walker (FLRW) metric, which is given by
\begin{equation}
    ds^2=-dt^2+a^2(t)\left[\frac{dr^2}{1-\kappa r^2}+r^2(d\theta^2+\sin^2\theta d\varphi^2)\right],
\end{equation}
where $a(t)$ is a scale factor that parametrizes the relative expansion of the universe and the curvature $\kappa$ is -1,0 or 1 when we are dealing respectively with an open, flat or closed universe.

\

Under the hypothesis that the FLRW spacetime is filled by a perfect fluid with energy density $\rho$ and pressure $P_T$, Einstein's field equations lead us to the following Friedmann equation
\begin{equation}
    H^2=\frac{\rho}{3}-\frac{\kappa}{a^2},
\end{equation}
where $H=\frac{\dot{a}}{a}$ is the Hubble parameter.

\

Note that this equation is only a constraint that relates the scale factor, the Hubble parameter and the energy density. Thus, in order to obtain the dynamical equations we will use the first principle of thermodynamics
$d(\rho V)=-P_TdV$, with $V=a^3$. Therefore, the conservation equation becomes
\begin{equation}
    \dot{\rho}=-3H(\rho+P_T)
    \label{cont}
\end{equation}
where the total pressure is $P_T=P+P_c$ with $P_c=-\frac{\Gamma}{3H}(\rho+P)$, being $P$ the pressure of the matter content and  $P_c$ the pressure related to the gravitationally induced adiabatic particle production with a creation rate $\Gamma$ (see for details \cite{SSL09,LJO10,LGPB14,NP15,LB10,LBC12,JOBL11}).

\

{\bf Remark}: Equation (\ref{cont}) could be obtained as follows:
Particle production is modeled by:
\begin{eqnarray}
\dot{N}=N\Gamma\Longleftrightarrow  \dot{n}+3Hn=n\Gamma,
\end{eqnarray}
where $N=na^3$ is the number of created particles and $n$ its density. Then, from Gibbs equation \cite{Prigogine88}
\begin{eqnarray}
nT \dot{s}=\dot{\rho}+3H\left(1-\frac{\Gamma}{3H}\right)(p+\rho),
\end{eqnarray}
where $T$ is the temperature of the fluid and $s$ the specific entropy (i.e., the entropy per particle), if one assumes that the specific entropy is constant (particle creation happens under adiabatic condition \cite{barrow90, lima92cov}), one obtains (\ref{cont}).

\

Differentiating Friedmann equation and using the conservation one, we derive the so-called Raychaudhuri equation
\begin{equation}
    \dot{H}=-\frac{\rho+P}{2}\left(1-\frac{\Gamma}{3H}\right)+\frac{\kappa}{a^2} .\label{raycht}
\end{equation}

From now on, we suppose that the perfect fluid satisfies a linear Equation of State (EoS) $P=(\gamma-1)\rho$, considering $\gamma>0$, which models a non-phantom fluid. The corresponding effective Equation of State, using \eqref{cont}, is given by
\begin{equation}
\omega_{\textit{eff}}=\frac{P-\frac{\Gamma}{3H}(\rho+P)}{\rho}=-1+\gamma\left(1-\frac{\Gamma}{3H}\right). \label{weff}
\end{equation}

In order to perform the dynamical analysis in both $\kappa=0$ and $\kappa=\pm 1$ cases, it will be useful to change to conformal time $\tau$, defined by the relation $d\tau=\frac{dt}{a}$, and to use the conformal Hubble parameter $\mathcal{H}=\frac{a'}{a}=\dot{a}$ where 
the prime denotes the derivative with respect to the conformal time and the dot with respect to the cosmic one. Hence, Friedmann equation and Raychaudhuri equations will become
\begin{equation}
    \mathcal{H}^2+\kappa=a^2\frac{\rho}{3} \ \ \ \ \ \
    \mathcal{H}'=-\frac{1}{2\mathcal{H}}[(3\gamma-2)\mathcal{H}-a\Gamma\gamma](\mathcal{H}^2+\kappa). \label{raychtau}
\end{equation}

Consequently, in the plane $({\mathcal H},a)$ the dynamical system is given by a two dimensional autonomous first order differential system:
\begin{eqnarray}\label{dynamical}
\left\{\begin{array}{ccc}
 \mathcal{H}' &=& - \frac{1}{2\mathcal{H}}[(3\gamma-2)\mathcal{H}-a\Gamma\gamma](\mathcal{H}^2+\kappa)\\
 a'&=& {\mathcal H} a
 \end{array}\right.,
 \end{eqnarray}
 and the energy density of the universe by
 \begin{equation}
 \rho=\frac{3}{a^2}( \mathcal{H}^2+\kappa).
 \end{equation}
 
\

\section{Constant production rate}

In this section we will study the case $\Gamma=\Gamma_0$ with $\Gamma_0>0$, whose dynamics was firstly introduced in the seminal paper \cite{Prigogine88}
and has recently been studied, for the flat case, in \cite{HP2016}.

\subsection{Flat universe}

For $\kappa=0$, equation \eqref{raycht} becomes
\begin{equation}
\dot{H}=-\frac{\gamma}{2}H(3H-\Gamma_0), \label{raychtflat}
\end{equation}
which means that,
in this case, the dynamical system decouples and we have a first order autonomous one dimensional dynamical system. Therefore, the dynamical analysis is very simple. We have two fixed points: $H=0$, which is a repeller, and $H=\frac{\Gamma_0}{3}$, which is an attractor, as we show in Figure \ref{rhoH}:

\begin{figure}[H]
\begin{center}
\includegraphics[height=35mm]{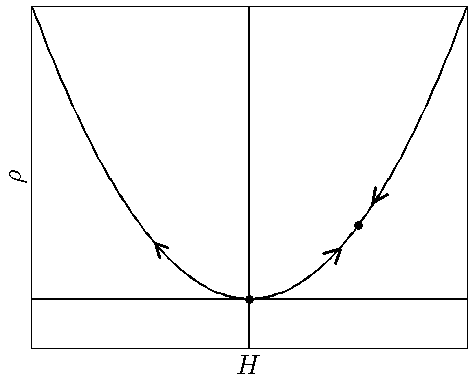}
\end{center}
\caption{Dynamics of the flat universe in the plane $(H,\rho)$. }
\label{rhoH}
\end{figure}

To study the singularities, we note that for large values of $|H|$ equation \eqref{raychtflat} becomes the same as in classical Einstein Cosmology, i.e., $\dot{H}=-\frac{3\gamma}{2}H^2$. Thus, we have a Big Bang singularity for $H>\frac{\Gamma_0}{3}$ and a Big Crunch singularity in the contracting phase.

\

Now, we will do the same dynamical analysis but in the plane $(\mathcal{H},a)$, because for  the case $\kappa\not=0$ it is mandatory to perform it in this plane. From equation \eqref{dynamical}, we obtain

\begin{equation}
\frac{d\mathcal{H}}{da}=-\frac{3\gamma-2}{2}\frac{\mathcal{H}}{a}+\frac{\Gamma_0\gamma}{2} \ \ \ \ \ \ \ \ \ \ \ \ \Longrightarrow \ \ \ \ \
\mathcal{H}=\frac{\Gamma_0}{3}a+Ca^{-\frac{3\gamma-2}{2}}.
\end{equation}

Consequently, the axis $\mathcal{H}=0$ is a repeller rect of critical points and, thus, we have no bounces. The axis $H=\frac{\Gamma_0}{3}\Longleftrightarrow\mathcal{H}=\frac{\Gamma_0}{3}a$, which corresponds to $\omega_{\textit{eff}}=-1$, is an attractor in the expanding phase, in the sense that all solutions satisfy $\frac{\mathcal{H}}{a}\rightarrow \frac{\Gamma_0}{3}$, at late times, i.e., the slope of the curves converges to $\frac{\Gamma_0}{3}$. When ${\mathcal H}\to\infty$, equation \eqref{raychtau} becomes ${\mathcal H}'=-\frac{1}{2}(3\gamma-2){\mathcal H}^2$. Therefore, as we have already explained, for cosmic time there is a Big Bang singularity when  $H>\frac{\Gamma_0}{3}$ and a Big Crunch in the expanding phase.

\

\begin{figure}[H]
\begin{center}
\includegraphics[height=55mm]{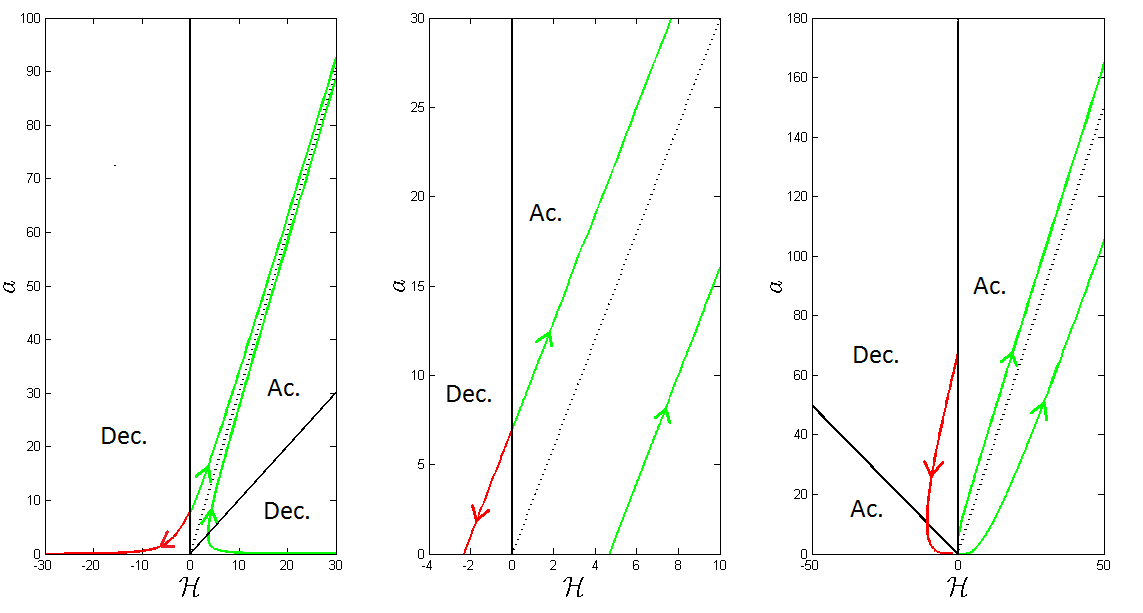}
\end{center}
\caption{Evolution of a flat universe ($\kappa=0$) in the plane $(\mathcal{H},a)$ with $\gamma>2/3$ (left), $\gamma=2/3$ (center) and $\gamma<2/3$ (right). In the accelerated region $\mathcal{H}'>0$, while in the decelerated region $\mathcal{H}'<0$.}
\label{figk0}
\end{figure}

In figure \ref{figk0} we have represented  the corresponding phase portraits where we have used colour green for the expanding phase and colour red for the contracting phase. Dotted lines correspond to the set $\omega_{\textit{eff}}=-1$. Black lines are the sets where $\mathcal{H}'=0$, i.e., $\mathcal{H}=0$ and, for $\gamma\neq 2/3$, $\mathcal{H}=a\frac{\Gamma_0\gamma}{3\gamma-2}$. We can see that for $\gamma>2/3$ there are solutions, in the expanding phase, with a Big Bang singularity that at early times depict a decelerating universe that at late time enter in an accelerated regime. On the contrary, for $\gamma\leq 2/3$, in the expanding phase, all the solutions depict, all the time, an accelerated universe.

\subsection{Closed universe}

For $\kappa\neq 0$, we cannot solve  analytically  the system for all the values of the parameter $ \gamma>0$. However, we can do it for the case ${\gamma=2/3}$. 
In this case, from \eqref{dynamical}
 we have  $\frac{d\mathcal{H}}{da}=\frac{\Gamma_0}{3}\frac{\mathcal{H}^2 +1}{\mathcal{H}^2}$ and, thus, orbits have the form ${\mathcal H}-\arctan{\mathcal H}=\frac{\Gamma_0}{3}a+C$. Therefore, all the orbits in the expanding phase, as in the flat case,  satisfy  ${\mathcal H}/a\rightarrow \Gamma_0/3$, meaning that $\omega_{\textit{eff}}\rightarrow -1$.

\

\begin{figure}[H]
\begin{center}
\includegraphics[height=40mm]{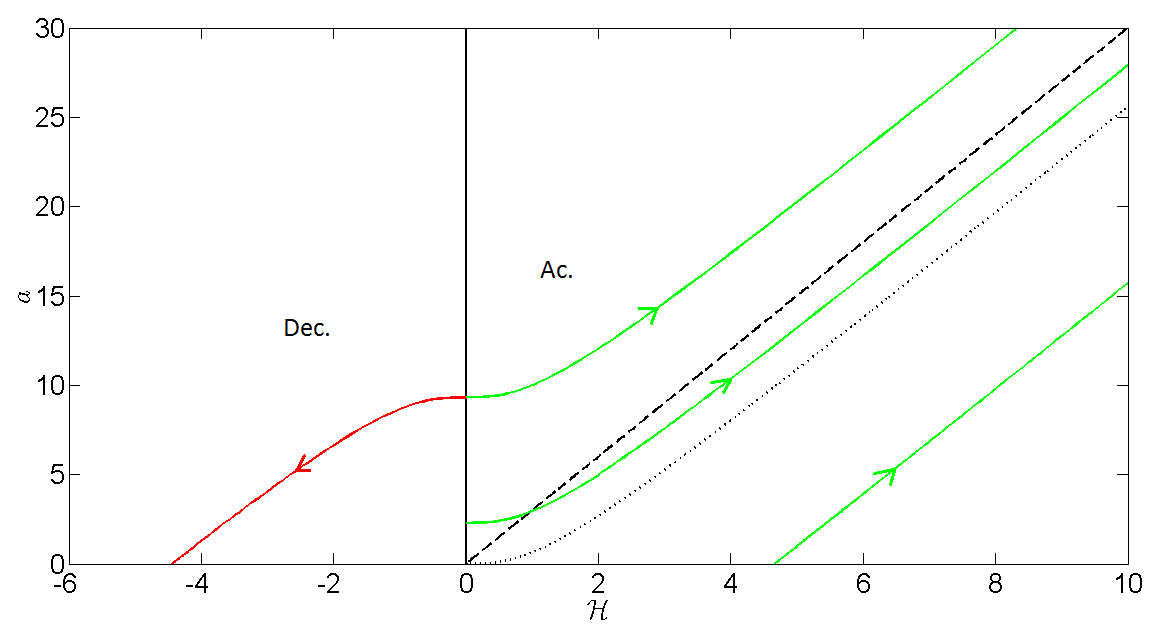}
\end{center}
\caption{Evolution of a closed universe ($\kappa=1$) in the plane $(\mathcal{H},a)$ with $\gamma=2/3$.}
\label{k1g23}
\end{figure}

In this case, the  orbit  $\mathcal{H}-\arctan\mathcal{H}=\frac{\Gamma_0}{3}a$, which corresponds to $C=0$, plays the role of the curve $
\mathcal{H}=\frac{\Gamma_0}{3}a \Longleftrightarrow \omega_{\textit{eff}}=-1$. However, since they differ for small values of $\mathcal{H}$, there are orbits which cross the curve $\omega_{\textit{eff}}=-1$, which means that for those orbits the universe goes from a phantom to a non-phantom regime.
The phase portrait for $\gamma=2/3$ is plotted in Figure \ref{k1g23}, where we have used the red colour for the contracting phase and the green colour for the expanding phase. The dotted line is the curve $C=0$ and the dashed line is the curve $\omega_{\textit{eff}}=-1$, which is an attractor in the same sense as in the flat case. Finally, ${\mathcal H}=0$ and $a=0$ (in the expanding phase) are repeller axis of singularities.

To show these singularities, 
let $\mathcal{H}_s$ be the value of $\mathcal{H}$ in the singularity corresponding to $a=0$, which is $0$ for the orbit with $C=0$ and finite otherwise. Hence, from $dt=da/\mathcal{H}$, we have that the time needed to reach the singularity is  $t=\int_0^{a_0}\frac{da}{\mathcal{H}}=\int_{\mathcal{H}_s}^{\mathcal{H}_0}\frac{3\mathcal{H}d{\mathcal H}}{\Gamma_0(\mathcal{H}^2+1)}$, which is finite. Hence, according to Friedmann equation ${\mathcal{H}}^2=\frac{\rho a^2}{3}-1$, the density of energy diverges for a finite cosmic time. Therefore, we have a Big Bang singularity for those orbits in the expanding phase such that $H-\frac{\arctan(aH)}{a}>\frac{\Gamma_0}{3}$ and a Big Crunch singularity in the contracting phase. 

\

In the same way,
we also observe some sort of singularity at ${\mathcal H}=0$. It occurs at finite time since $t=\int_0^{\mathcal{H}_0}\frac{3\mathcal{H}d{\mathcal H}}{\Gamma_0(\mathcal{H}^2+1)}$ does not diverge. Hence, ${\mathcal H}'$ diverges at ${\mathcal H}=0$, which means that the scalar curvature $R=\frac{6}{a^2}({\mathcal H}'+ {\mathcal H}^2+\kappa)$ diverges at finite time.

%$a\to a_s>0$ and, from Friedmann equation, $\rho\to\rho_s$. We see, as well, that, given that $H\to 0$, $|P_T|\to\infty$. Therefore, both in the contracting phase and in the orbits in the expanding phase such that $H-\frac{\arctan(aH)}{a}<\frac{\Gamma_0}{3}$, there is a Past Sudden singularity. 

\

For $\mathbf{\gamma\neq 2/3}$, we also have another orbit, that we have represented in a dotted line, which plays the role of the curve $\omega_{\textit{eff}}=-1$, i.e, $\frac{\mathcal{H}}{a}=\frac{\Gamma_0}{3}$ (dashed line). The other black line refers to $\frac{\mathcal{H}}{a}=\frac{\Gamma_0\gamma}{3\gamma-2}$, where $\mathcal{H}'=0$. And the axis $\mathcal{H}=0$ is a repeller rect of singularities (the scalar curvature diverges on it).

\begin{figure}[H]
\begin{center}
\includegraphics[height=50mm]{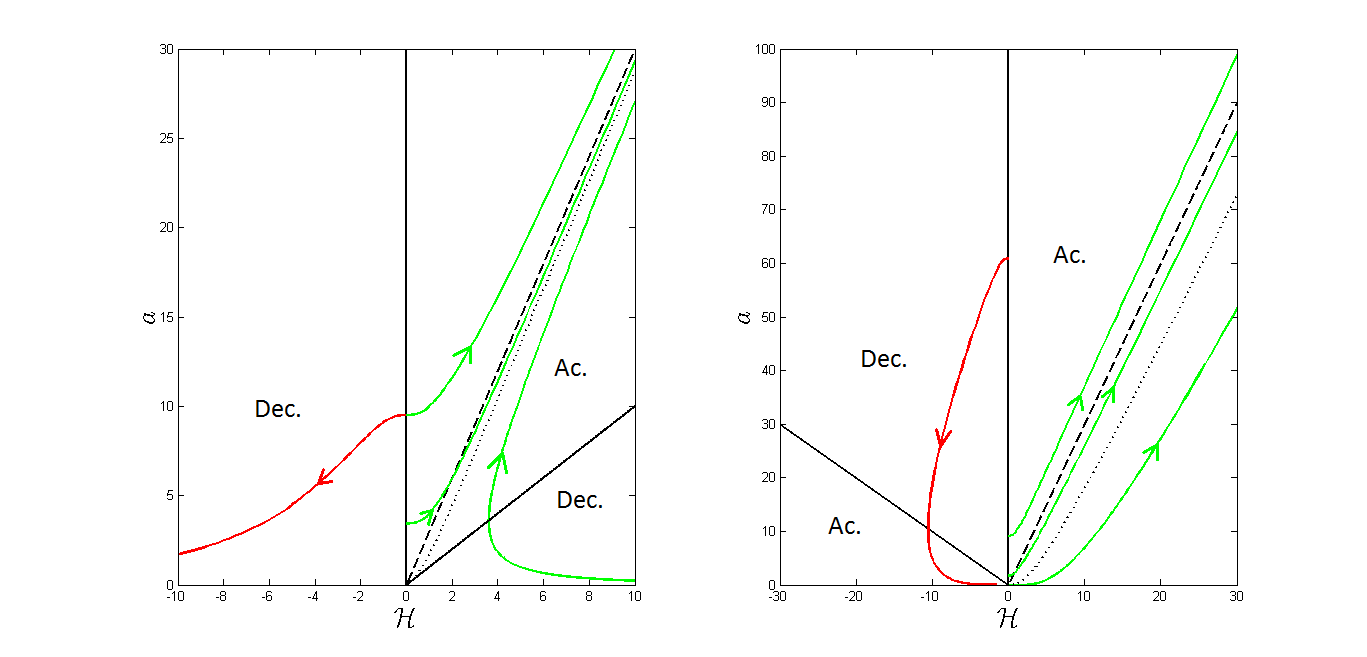}
\end{center}
\caption{Evolution of a closed universe ($\kappa=1$) in the plane $(\mathcal{H},a)$ with $\gamma>2/3$ (left) and $\gamma<2/3$ (right).}
\end{figure}

We observe that the behavior in terms of singularities is the same as the one for $\gamma=2/3$. We only need to verify that they occur for a finite cosmic time. Regarding the Big Bang and Big Crunch singularity, it is trivial since in these cases when ${\mathcal H}\to \pm\infty$  the dynamics is the same as  for $\kappa=0$. Finally, near $\mathcal{H}=0$, equation \eqref{dynamical} becomes $\mathcal{H}'=\frac{a\Gamma\gamma}{2\mathcal{H}}$, and hence, the  time to reach ${\mathcal H}=0$, i.e. $R=\infty$,  does not diverge.

\

In the phase portrait, we have also plotted an orbit that crosses the line $\omega_{\textit{eff}}=-1$, i.e., going from a phantom to a non-phantom phase.
Anyway, in the expanding phase all the orbits satisfy  $\frac{\mathcal{H}}{a}\rightarrow \frac{\Gamma_0}{3}\Longrightarrow \omega_{\textit{eff}}\rightarrow -1$, and as in the flat case,  when $\gamma>2/3$ there are orbits  depicting, at early times,  a decelerating universe  that accelerates at late times.

\subsection{Open universe}

\begin{figure}[H]
\begin{center}
\includegraphics[height=45mm]{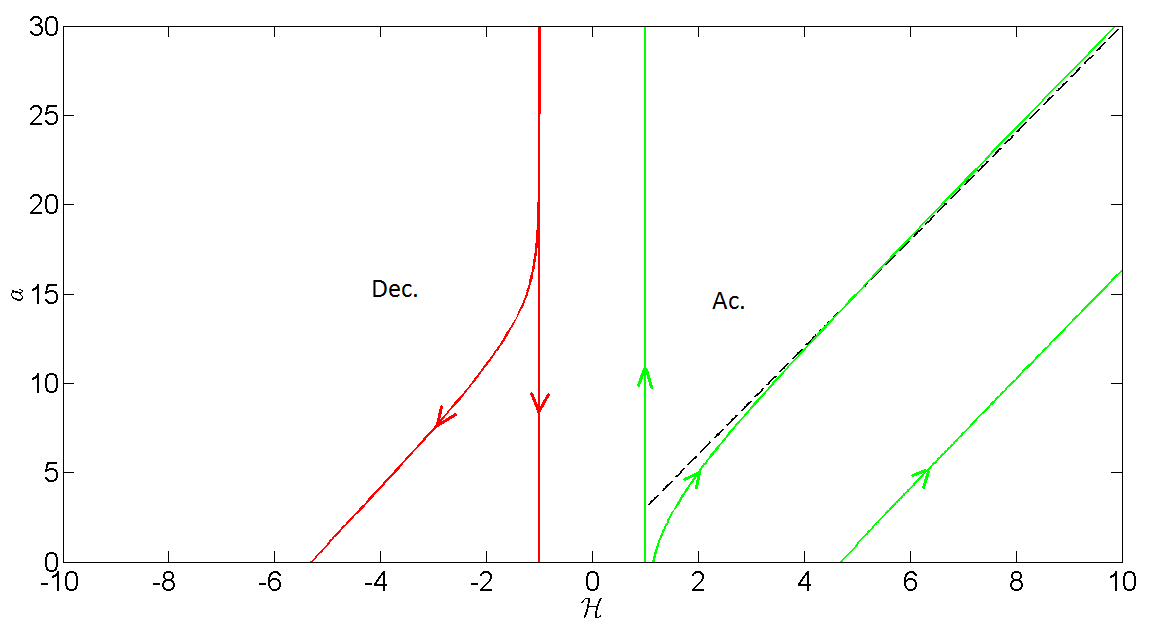}
\end{center}
\caption{Evolution of an open universe ($\kappa=-1$) in the plane $(\mathcal{H},a)$ with $\gamma=2/3$.}
\label{k-1}
\end{figure}

Analogously as in the closed case, we obtain for $\mathbf{\gamma=2/3}$ the equation $\frac{d\mathcal{H}}{da}=\frac{\Gamma_0}{3}\frac{\mathcal{H}^2-1}{\mathcal{H}^2}$, which could be integrated obtaining  orbits of the form
$\mathcal{H}+\frac{1}{2}\ln\frac{\mathcal{H}-1}{\mathcal{H}+1}=\frac{\Gamma_0}{3}a+C$. Therefore, all the orbits in the expanding phase satisfy that ${\mathcal H}/a\rightarrow \Gamma_0/3$ and, thus, $\omega_{\textit{eff}}\rightarrow -1$. We note that in this case, from the Friedmann equation  $\rho=\frac{3}{a^2}( \mathcal{H}^2-1)$, one can deduce that the  region $|\mathcal{H}|<1$ is forbidden because  the energy density $\rho$ must be positive. The  behavior of the orbits is depicted in Figure \ref{k-1}.

To study the singularities, 
let $\mathcal{H}_s$ be the value of $\mathcal{H}$ in the singularity corresponding to $a=0$, which is a finite value satisfying $|\mathcal{H}_s|>1$. Hence, from $dt=da/\mathcal{H}$, we deduce that the time needed to reach the singularity is $t=\int_0^{a_0}\frac{da}{\mathcal{H}}=\int_{\mathcal{H}_s}^{\mathcal{H}_0}\frac{3\mathcal{H}d\mathcal{H}}{\Gamma_0(\mathcal{H}^2-1)}$, which is finite. Hence, according to Friedmann equation ${\mathcal{H}}^2=\frac{\rho a^2}{3}+1$, the density of energy diverges for a finite cosmic time. Therefore, we have a Big Bang singularity for orbits in the expanding phase and a Big Crunch singularity in the contracting phase. 

\

The phase portraits in the case $\mathbf{\gamma\neq 2/3}$ are:

\begin{figure}[H]
\begin{center}
\includegraphics[height=50mm]{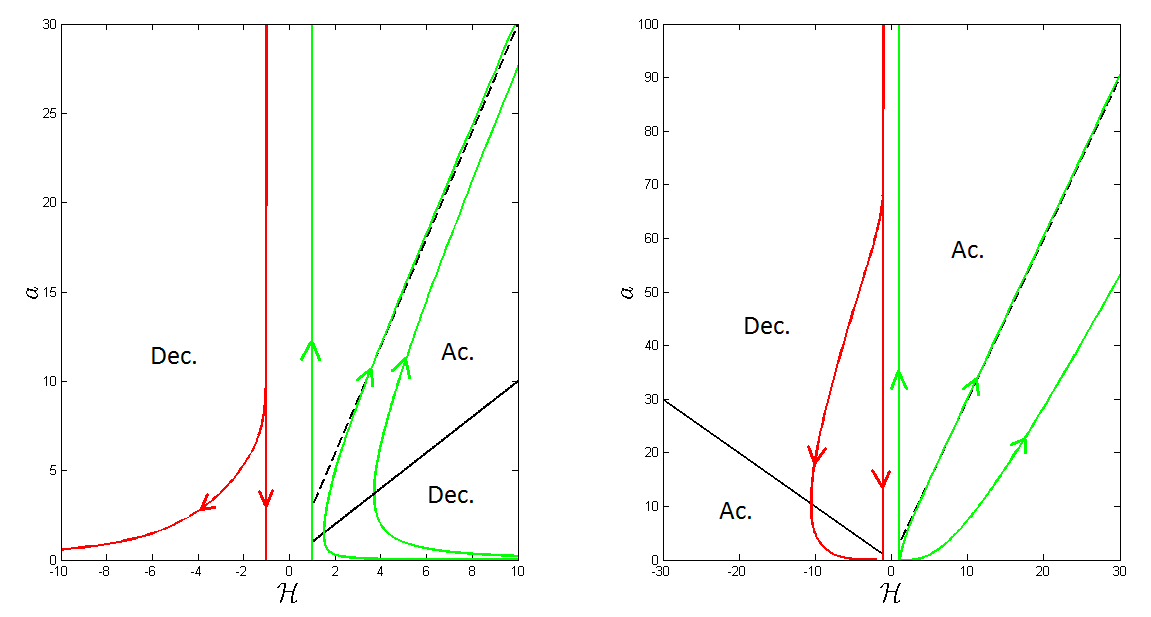}
\end{center}
\caption{Evolution of an open universe ($\kappa=-1$) in the plane $(\mathcal{H},a)$ with $\gamma>2/3$ (left) and $\gamma<2/3$ (right).}
\end{figure}

We observe that the behavior in terms of singularities is the same as  for $\gamma=2/3$. We only need to verify that they occur for a finite cosmic time. With the same argument  used in the closed case, the Big Bang and Big Crunch singularity occur at a finite time because in these cases $|\mathcal {H}|\to\infty$ and, thus, it satisfies asymptotically the same dynamics as for $\kappa=0$.

\

In the phase portrait, we have also plotted an orbit that crosses the line $\omega_{\textit{eff}}=-1$. In these cases, the orbits $\mathcal{H}=\pm 1$ (i.e. $\rho=0$) are repellers, while in the expanding phase all the orbits satisfy as in the cases $\kappa=0, 1$ that $H\rightarrow \frac{\Gamma_0}{3}$, i.e., $\omega_{eff}\rightarrow -1$.

\

\section[]{Variable production rate}

In \cite{PHPS2016}, the authors considered a variable production rate given by $\Gamma(H)=-\Gamma_0+mH+\frac{n}{H}$ and studied the different possible values of the constants. It is discussed that for some particular cases it is possible to depict, for the flat case, a non-singular universe in cosmic time, i.e., defined for all $t\in (-\infty,\infty)$,  with an early accelerated period (inflation), that decelerates after the end of inflation and finally enters in another accelerated regime which could mimic the current cosmic acceleration. In this section we are going to consider one of this cases, which results when $m=3$, $\Gamma_0>0$ and $n>0$.

%In this section we will study  a production rate given by $\Gamma(H)=-\Gamma_0+3H+\frac{n}{H}$, $\Gamma_0>0$, $n>0$. As has been  recently showed  in \cite{PHPS2016}, it depicts, for the flat case, a universe with an early accelerated period (inflation), that decelerates after the end of inflation and finally enters in another accelerated regime which could mimic the current cosmic acceleration.

\

 The viability of this phenomenological model was proved in \cite{PHPS2016} for a flat universe  in the particular case in which $n=\frac{\Gamma_0^2}{12}$, verifying that the spectral index ($n_s$), its running ($\alpha_s$) and the ratio of tensor to scalar perturbations (r) fit well with the observed values.

\subsection{Flat universe}

For $\kappa=0$, equation \eqref{raycht} becomes:
\begin{equation}
\dot{H}=-\frac{\gamma}{2}(\Gamma_0H-n) \label{raychflatG}
\end{equation}

So, we have a single fixed point, $H=\frac{n}{\Gamma_0}$, which is an attractor:
\begin{figure}[H]
\begin{center}
\includegraphics[height=30mm]{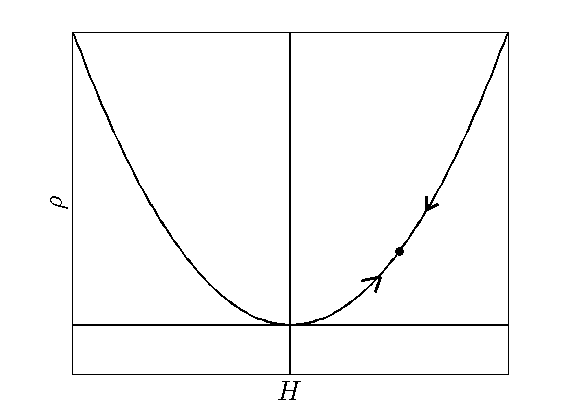}
\end{center}
\caption{Dynamics of the flat universe in the plane $(H,\rho)$ }
\end{figure}

Since for $H\to\infty$ equation \eqref{raychflatG} is $\dot{H}=-\frac{\gamma\Gamma_0}{2}H$,  we deduce that time to reach $H\to\pm\infty$ diverges and, therefore, we have no singularities in {\color{red} finite} cosmic time. We also observe that for $H<\frac{n}{\Gamma_0}$ there is a bounce.

\

As in the former case, we will perform the dynamical analysis in the plane $(\mathcal{H},a)$.  
The equation (\ref{raychflatG}) could be written as
$\frac{dH}{d\ln a}=\frac{\gamma}{2}\left( \frac{n}{H}-\Gamma_0 \right)$,
whose general solution is given by
%From $a'=\mathcal{H}{a}$, we obtain that
\begin{equation}
 \ln a=C-\frac{2}{\gamma\Gamma_0}\left(\frac{\mathcal{H}}{a}+\frac{n}{\Gamma_0}\ln\left|\frac{\mathcal{H}}{a}-\frac{n}{\Gamma_0} \right|\right).
\end{equation}

Moreover, from equation $\frac{dH}{d\ln a}=\frac{\gamma}{2}\left( \frac{n}{H}-\Gamma_0 \right)$, we observe that all orbits approach asymptotically to $\frac{\mathcal{H}}{a}=\frac{n}{\Gamma_0}$, i.e., $\omega_{\textit{eff}}\to-1$. We point out that $\mathcal{H}'=0$ when

\begin{equation}
\frac{a}{\mathcal{H}}=\frac{\gamma\Gamma_0\pm\sqrt{\gamma^2\Gamma_0^2-8n\gamma}}{2n\gamma}
\end{equation}

Hence, we differ between whether these two rects in the plane $(\mathcal{H},a)$ are complex ($\gamma\Gamma_0^2-8n<0$), coincide ($\gamma\Gamma_0^2=8n$) or are both real and different ($\gamma\Gamma_0^2-8n>0$). In the following phase portraits we can verify the behaviour that we have already described:

\begin{figure}[H]
\begin{center}
\includegraphics[height=55mm]{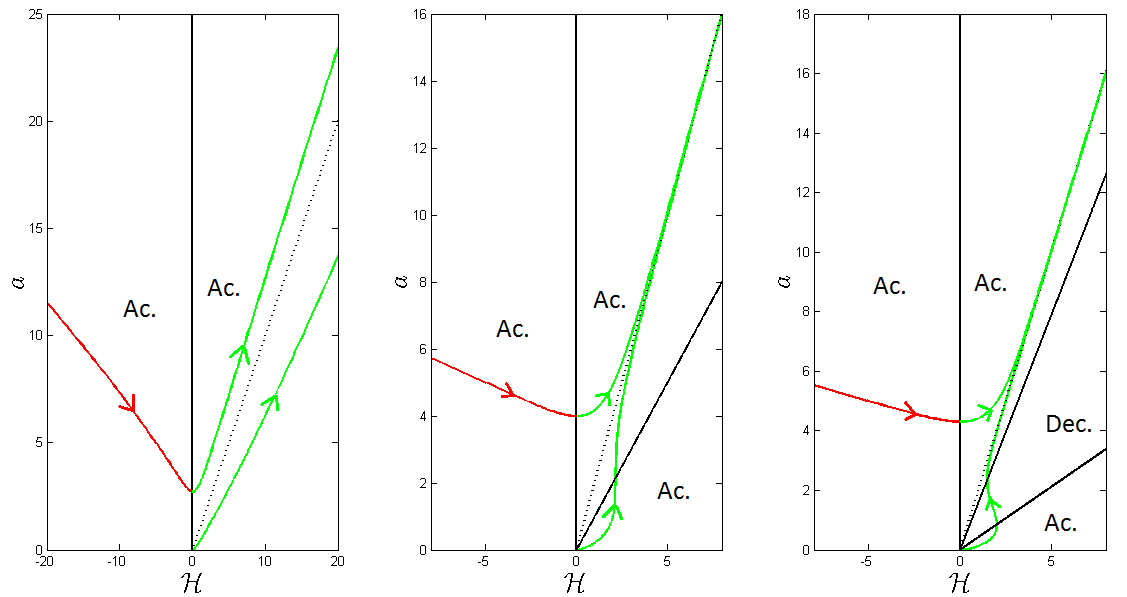}
\end{center}
\caption{Evolution of a flat universe ($\kappa=0$) in the plane $(\mathcal{H},a)$ with $\gamma\Gamma_0^2-8n<0$ (left), $\gamma\Gamma_0^2-8n=0$ (center) or $\gamma\Gamma_0^2-8n>0$ (right).}
\end{figure}

\

We note that in the case $\gamma\Gamma_0^2-8n>0$, there are orbits that depict a universe accelerating at early times and late times, with a deceleration period between these accelerated phases. Then, these orbits could be candidates to unify inflation with the current cosmic acceleration. We also point out that these orbits have no singularities in finite cosmic time, since at early times $H(t)\to \frac{2n\gamma}{\gamma\Gamma_0-\sqrt{\gamma^2\Gamma_0^2-8n\gamma}}$ and, thus, point $(0,0)$ in the plane $(\mathcal{H},a)$ is not reached in a finite time, which corresponds to the so-called \emph{little bang} singularity \cite{inflation1}, named in analogy to the \emph{little rip} \cite{frampton,nojiri}.

\

However, we see that the background is not past-complete analogously to the analysis done in \cite{inflation1}. This follows from the fact that the maximum affine parameter $\tilde{\lambda}_{\textit{max}}=\frac{1}{a_0}\int_{-\infty}^0a(t)dt$ is finite. Given that at sufficiently early times $\dot{a}\simeq\frac{2n\gamma}{\gamma\Gamma_0-\sqrt{\gamma^2\Gamma_0^2-8n\gamma}}a$ (let us consider for $t<t_c$), $\int_{-\infty}^{t_c}a(t)\simeq\frac{\gamma\Gamma_0-\sqrt{\gamma^2\Gamma_0^2-8n\gamma}}{2n\gamma}e^{\frac{2n\gamma t_c}{\gamma\Gamma_0-\sqrt{\gamma^2\Gamma_0^2-8n\gamma}}}<\infty$ and, thus, $\tilde{\lambda}_{\textit{max}}$ is finite, i .e., any backward-going null geodesic has a finite affine length.

\

With regards to massive particles moving along time-like geodesics, being $m$ its mass and {$p_0\not=0$} its three-momentum at time $t=0$, we verify that the maximum proper time $\tau_{\textit{max}}\equiv\int_{-\infty}^0\frac{ma(t)}{\sqrt{m^2a^2(t)+p_0^2a^2(0)}}dt$ will also be finite.  { Hence,  in the system of reference that a massive particle
moving along a time-like geodesic is at rest, the
singularity (the divergence of the energy density) 
occurs at finite proper time \cite{leonardo}, meaning that the universe is not past-complete.   Note finally that,  as has been pointed out in for the first time in the seminal paper \cite{bvg},
this past incompleteness is a an universal property 
 of the scenarios that contains an early inflationary period.}

\

\subsection{Closed universe}

The dynamical system associated to this case is:
\begin{equation}
    \mathcal{H}'=\left(\mathcal{H}^2-\frac{\gamma\Gamma_0}{2}a\mathcal{H}+\frac{n\gamma}{2}a^2\right)\frac{\mathcal{H}^2+1}{\mathcal{H}^2}.
\end{equation}

In this case, $\mathcal{H}=0$ is a repeller axis of singularities and, hence, we have no bounce. Since for $\mathcal{H}\to\pm\infty$ the behavior is the same as in the flat case, the time needed to reach $\mathcal{H}\to\pm\infty$ will diverge again. The phase portrait for the different cases, which we have already described for the flat case,  are the following ones:

\begin{figure}[H]
\begin{center}
\includegraphics[height=55mm]{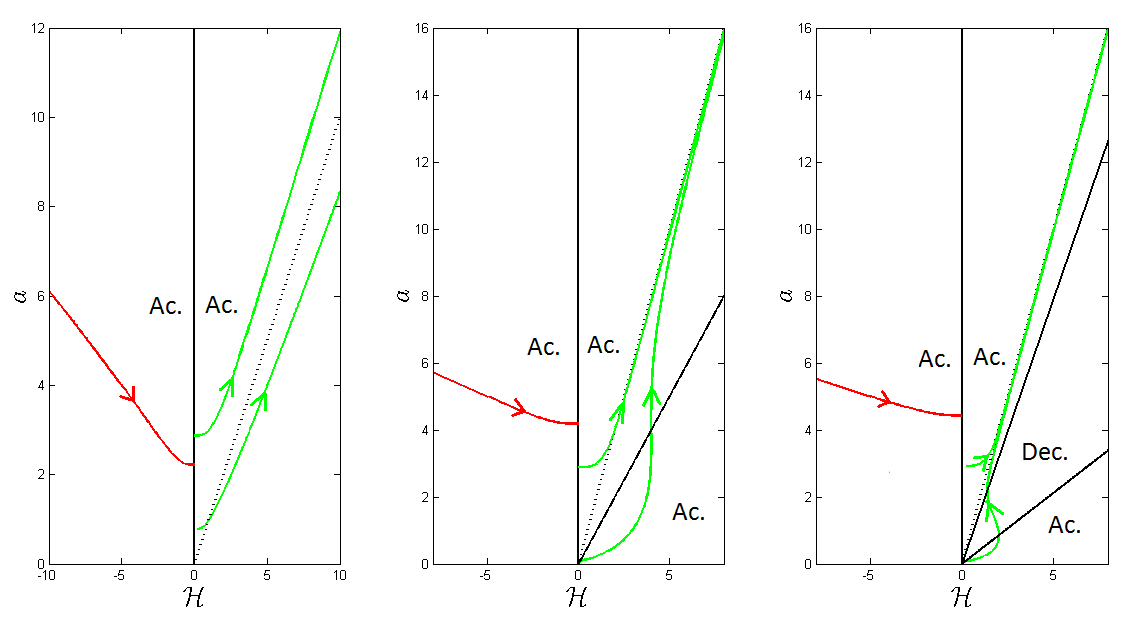}
\end{center}
\caption{Evolution of a closed universe ($\kappa=1$) in the plane $(\mathcal{H},a)$ with $\gamma\Gamma_0^2-8n<0$ (left), $\gamma\Gamma_0^2-8n=0$ (center) or $\gamma\Gamma_0^2-8n>0$ (right).}
\end{figure}

All orbits in the contracting phase contain a finite time future singularity because near $\mathcal{H}=0$ the dynamical equation becomes $\mathcal{H}'=\frac{n\gamma}{2}\frac{a^2}{\mathcal{H}^2}$, which shows that the derivative of $\mathcal{H}$ diverges (and, thus, the scalar curvature diverges). Moreover, it is not difficult to see that the time to reach $\mathcal{H}=0$ is finite.

\

In the expanding phase, $H=\frac{n}{\Gamma_0}$ is an attractor. Some orbits have a past singularity in which $\mathcal{H}'$ diverges, as in the contracting phase. When $\gamma\Gamma_0^2-8n> 0$, analogously as in the flat case, we also have orbits without singularities in finite cosmic time which could depict the unification of the early inflationary period with the late time cosmic acceleration.

%that tend for $t\to-\infty$ to $H=\frac{2n\gamma}{\gamma\Gamma_0-\sqrt{\gamma^2\Gamma_0^2-8n\gamma}}$ ($\mathcal{H}\to 0$ and $a\to 0$).

\subsection{Open universe}

The corresponding dynamical system is:
\begin{equation}\label{x}
    \mathcal{H}'=\left(\mathcal{H}^2-\frac{\gamma\Gamma_0}{2}a\mathcal{H}+\frac{n\gamma}{2}a^2\right)\frac{\mathcal{H}^2-1}{\mathcal{H}^2}
\end{equation}

\

As in the closed universe, for $\mathcal{H}\to\pm\infty$ we have no singularities in cosmic time. And the axis $\mathcal{H}=\pm 1$ represents solutions of the dynamical system. The phase portrait is:

\begin{figure}[H]
\begin{center}
\includegraphics[height=55mm]{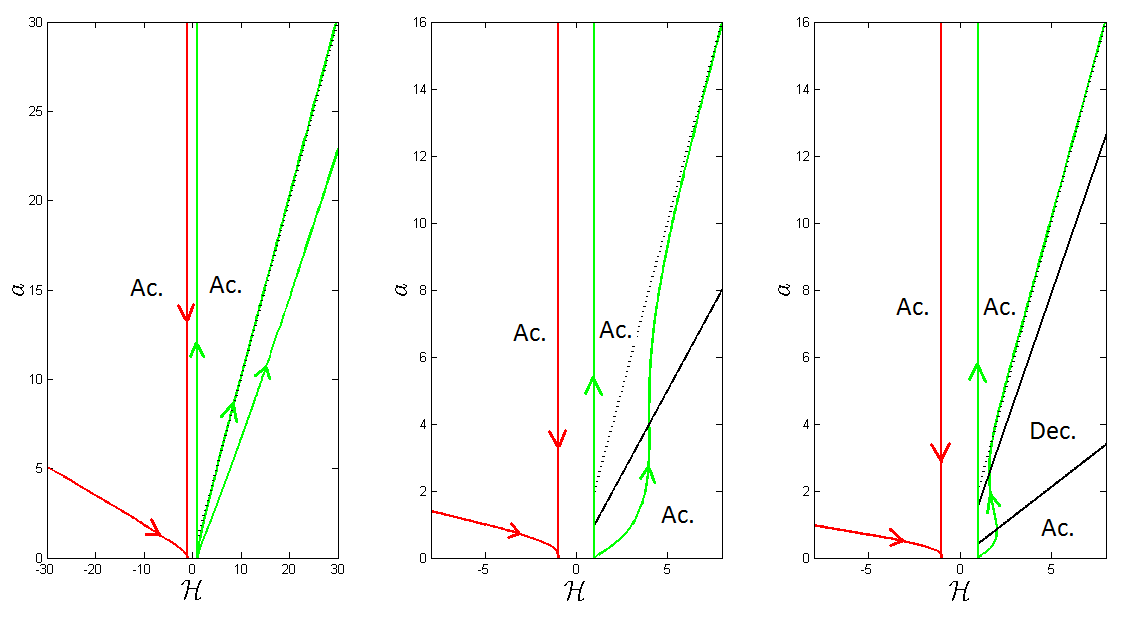}
\end{center}
\caption{Evolution of an open universe ($\kappa=-1$) in the plane $(\mathcal{H},a)$ with $\gamma\Gamma_0^2-8n<0$ (left), $\gamma\Gamma_0^2-8n=0$ (center) or $\gamma\Gamma_0^2-8n>0$ (right).}
\end{figure}

\

Since $\dot{a}=\mathcal{H}$, the time needed to reach the point $(\pm 1,0)$ is finite. Therefore, $H$ diverges at finite cosmic time, but this does no mean that there is a singularity. Effectively, near $(\pm 1,0)$ the dynamical equation (\ref{x}) becomes 
\begin{equation}
 \mathcal{H}'=\mathcal{H}^2-1 \Longleftrightarrow \frac{d\mathcal{H}}{da} =\frac{\mathcal{H}^2-1}{a\mathcal{H}},
 \end{equation}
whose general solution is $\mathcal{H}^2-1=(\mathcal{H}^2_0-1)\left(\frac{a}{a_0} \right)^2$. 
Then, since $\rho=\frac{3}{a^2}(\mathcal{H}^2-1)$ and near $(\pm 1,0)$ the scalar curvature is approximately $2\rho$, we can conclude that both quantities are finite at $(\pm 1,0)$. Finally, once again, in the expanding phase $H=\frac{n}{\Gamma_0}$ is an attractor and there are orbits that could be candidates to depict our universe.

\

\section{Thermodynamical analysis}

This section is devoted to the  thermodynamical study  of the  models presented in the previous sections. Macroscopic systems tend toward a thermodynamical equilibrium where, according to the second law of thermodynamics,  the total entropy  $S=S_h+S_{\gamma}$ (the  entropy of the apparent horizon, namely $S_h$, plus the entropy of matter enclosed by the horizon, that we denote by $S_{\gamma}$) of an isolated system never decreases, that is $\dot{S}_h+ \dot{S}_{\gamma} \geq  0$, and  should be concave ($\ddot{S}_h+ \ddot{S}_{\gamma} <0$) near the thermodynamic equilibrium \cite{MP13, pavon1}.
Here, we will use units $\hbar=c=k_B=1$ and now the reduced Planck's mass is $M_{Pl}=\frac{1}{\sqrt{8\pi G}}$.

%and $\ddot{S} < 0$, that means it should be concave at least in the last stage of approaching thermodynamic equilibrium \cite{MP13}.

%The total entropy  (the  entropy of the apparent horizon plus the matter or any fields enclosed by the horizon) should satisfy $\dot{S}_h+ \dot{S}_{\gamma} \geq  0$, where $S_h$ stands for the entropy of the apparent horizon and $S_{\gamma}$ for the matter fields. On the other hand, $\ddot{S}_h+ \ddot{S}_{\gamma}  < 0$   at very late time. 

%The entropy of the apparent horizon is defined as $S_h = k_{B} \mathcal{A}/4\, l_{pl}^2$, where $k_{B}$ is the Boltzmann's constant, $\mathcal{A}= 4 \pi r_h^2$, is the horizon area in which $r_h= \frac{1}{H}$ being the Hubble radius \cite{Bak}, and $l_{pl}=\sqrt{\frac{1}{8\pi}}$ is the Planck's length in the units used in the present work.

% However, we mention that for the present work, we have considered the spatial flatness of the FLRW universe, that means, in our case, the horizon radius reduces to $r_h = 1/H$.

%\begin{eqnarray}\label{newsp1}
%H = \frac{\Gamma}{3}+ \left(H_0 -\frac{\Gamma}{3} \right) \left(\frac{a_0}{a}\right)^{3\gamma/2}
%\end{eqnarray}
%where $a_0$, $H_0$ are respectively the current values of the scale factor and the Hubble parameter. Considering the entropy of the apparent horizon, we find that

%\begin{eqnarray}
%S_h^\prime &=& - \left(\frac{2\,k_B\, \pi}{l_{pl}^2}\right)\, \frac{H^\prime}{H^3}\\
%&=&  \left(\frac{2\,k_B\, \pi}{l_{pl}^2\, %H^3}\right)\, \left(H_0 - \frac{\Gamma}{3} %\right)\, a_0^{\frac{3\gamma}{2}}\, %a^{-\frac{3\gamma}{2} -1}
%\end{eqnarray}

\subsection[]{Constant creation rate ($\Gamma=\Gamma_0$)}

The entropy of the apparent horizon is defined as $S_h=\frac{1}{4l_{\textit{Pl}}^2}\mathcal{A}$, where $\mathcal{A}=4\pi r_h^2$ is the horizon area, $r_h=\frac{1}{\sqrt{H^2+\frac{k}{a^2}}}=\sqrt{\frac{3}{\rho}}$ is the horizon radius and $l_{\textit{Pl}}=\frac{1}{\sqrt{8\pi}M_{Pl}}$ is Planck's length. Now, a straightforward calculus leads  to
\begin{equation}
\dot{S}_h=\frac{3\gamma}{4} \frac{\mathcal A}{l_{Pl}^2}\left(H-\frac{\Gamma_0}{3}\right),
\end{equation}
which shows that $\dot{S}_h>0$ for  the orbits that satisfy $H>\frac{\Gamma}{3}$ ($\omega_{eff}>-1$), and of course the entropy of the apparent horizon decreases in the contacting phase, in agreement with the recent result obtained
in \cite{pavon}. Taking the second derivative, we obtain 
\begin{equation}
\ddot{S}_h=  \frac{3\gamma}{4} \frac{\mathcal A}{l_{Pl}^2}\    \left[\frac{3\gamma}{2}\left(1-\frac{\Gamma_0}{3H} \right)\left(H^2\left(1-\frac{2\Gamma_0}{3H}\right) -\frac{\kappa}{a^2} \right)+\frac{\kappa}{a^2} \right].
\end{equation}

When the system tends to the thermodynamical equilibrium $H=\frac{\Gamma_0}{3}$, we have seen that $a\to\infty$ and, thus, it is trivial that for $\kappa=0,-1$ it is concave ($\ddot{S}_h<0$) in the last stage of approaching this equilibrium. However, for $\kappa=1$, the study is a little more involved. First of all, we note that near the equilibrium one has
\begin{equation}
\ddot{S}_h=  \frac{3\gamma}{4} \frac{\mathcal A}{l_{Pl}^2 a^2}\ \left[-\frac{\gamma\Gamma_0^2}{6}f+1 \right],
\label{ddshk1}
\end{equation}
where we  have defined $f=a^2\left(1-\frac{\Gamma_0}{3H}\right)>0$.

%we have to study whether $\frac{3\gamma}{2}H^2a^2\left(1-\frac{\Gamma_0}{3H}\right)$ is greater or less than $1$.

\

%Let $f(t):=a(t)^2\left(1-\frac{\Gamma_0}{3H(t)}\right)$. 
Then, we see that near $H=\frac{\Gamma_0}{3}$, $f$ satisfies
\begin{equation}
\dot{f}=\frac{\Gamma_0(4-3\gamma)}{6}f+\frac{3}{\Gamma_0}\qquad \Longrightarrow \qquad  f(t)= Ce^{\frac{\Gamma_0 (4-3\gamma)}{6}t}+\frac{18}{\Gamma_0^2(3\gamma-4)}.
\end{equation}

Hence, $\lim\limits_{t\to\infty}f(t)=\infty$ if $4-3\gamma\geq0$ and $\lim\limits_{t\to\infty}f(t)=\frac{18}{\Gamma_0^2(3\gamma-4)}$ if $4-3\gamma<0$. Therefore, if $4-3\gamma\geq0$ we easily verify that for the closed case $\ddot{S}_h$ is concave in the last stage of approaching the equilibrium. If $4-3\gamma<0$, we also obtain concavity since near $H=\frac{\Gamma_0}{3}$, it is satisfied that
$-\frac{\gamma\Gamma_0^2}{6}f+1\rightarrow \frac{4}{4-3\gamma}<0$.

\

Now, we need to study the entropy $S_{\gamma}$ of the field, which arises from Gibb's equation $TdS_{\gamma}=d(\rho V)+PdV$, where $V=\frac{4}{3}\pi r_h^3$. Thus,
\begin{equation}
T\dot{S}_{\gamma}={6\gamma\pi M_{Pl}^2r_{h}} \left(H-\frac{\Gamma}{3}\right)(3\gamma-2),
\end{equation}
which is positive in the case we study ($H>\frac{\Gamma_0}{3}$)  when $\gamma>\frac{2}{3}$. On the other hand, from the equation \cite{pavon1,Zimdahl2}
\begin{equation}
\frac{\dot{T}}{T}=(\Gamma-3H)\frac{\partial P}{\partial\rho},
\end{equation}
we obtain, using $\Gamma-3H=\frac{2H\left(\dot{H}-\frac{\kappa}{a^2}\right)}{\gamma\left(H^2+\frac{\kappa}{a^2}\right)}=-\frac{2}{\gamma}\frac{d}{dt}\ln r_h$, the general solution
\begin{equation}
T=T_0\left(\frac{r_{h,0}}{r_h}\right)^{\frac{2(\gamma-1)}{\gamma}},
\end{equation}
where we have introduced the  notation $r_{h,0}=\frac{1}{\sqrt{H_0^2+\frac{k}{a_0^2}}}$

Regarding the case $0<\gamma<2/3$, we need to look more carefully at the sign of the derivative of the total entropy.

\begin{equation}
\dot{S}=\frac{6\pi\gamma\left(H-\frac{\Gamma_0}{3}\right)M_{Pl}^2}{\sqrt{H^2+\frac{\kappa}{a^2}}}\left[\frac{4\pi}{\sqrt{H^2+\frac{\kappa}{a^2}}}-\frac{2-3\gamma}{T}\right]
\label{dstot}
\end{equation}

According to 2015 Planck results \cite{Planck2015}, $|\Omega_k|<0.005$ and, hence, $H_0^2a_0^2>200$. Thus,  $\frac{k}{a_0^2}\ll H_0^2$, 
%since $H_0\cong 6\times10^{-61}M_{Pl}$, one has $a_0^2>\frac{200}{H_0^2}\cong  10^{124} M_{Pl}^{-2}$ 
meaning that $r_{h,0}\approx \frac{1}{H_0}$ and, since we are currently in the last accelerated phase expansion phase of the universe, $H_0\approx\frac{\Gamma_0}{3}$. Therefore, at the last stage of approaching equilibrium $T\approx T_0$. Hence, $\frac{4\pi T}{(2-3\gamma)\sqrt{H^2+\frac{\kappa}{a^2}}}>\frac{2T_0}{H_0}$, which is of the order $10^{-30}$  because the current temperature of the universe is $T_0\cong10^{-31}M_{Pl}$ and the current value of the Hubble parameter is $H_0\cong 6\times10^{-61}M_{Pl}$. So, it turns out that for $0<\gamma<2/3$ it is also fulfilled that $\dot{S}>0$.

\

Now, the second derivative of $S_{\gamma}$ is given by
\begin{equation}
\ddot{S}_{\gamma}=\left\{\begin{array}{ll}
        \frac{18\pi\gamma(3\gamma-2)(\gamma-1)M_{Pl}^2r_h}{T}\left\{\left(1-\frac{\Gamma_0}{3H}\right)\left[H^2\left(1-\frac{3\gamma-2}{6(\gamma-1)}\frac{\Gamma_0}{H}\right)-\frac{\gamma}{2(\gamma-1)}\frac{\kappa}{a^2} \right]+\frac{1}{3(\gamma-1)}\frac{\kappa}{a^2}\right\},  & \mbox{$\gamma\neq1$}\\
        \frac{6\pi\gamma(3\gamma-2)M_{Pl}^2r_h}{T}\left[-\frac{1-\frac{\Gamma_0}{3H}}{2}\left(H\Gamma_0+3\frac{\kappa}{a^2}\right)+\frac{\kappa}{a^2}\right] & \mbox{$\gamma=1$}.\end{array}\right.
\end{equation}

Both for $\gamma=1$ and the rest of cases such that $\gamma>2/3$, we trivially see that $\ddot{S}_{\gamma}<0$ in the last stage of approaching equilibrium for $\kappa=0$ and $\kappa=-1$, 
because  $\frac{3\gamma-2}{6(\gamma-1)}\frac{\Gamma_0}{H}>1$ for $\gamma>1$ and $\frac{3\gamma-2}{6(\gamma-1)}<0$ for $\gamma<1$. For $\kappa=1$, we need to study again the function $f=a^2\left(1-\frac{\Gamma_0}{3H}\right)$  near $H=\frac{\Gamma_0}{3}$. Performing exactly the same analysis as the one we did for $\ddot{S}_h$, one can conclude that for the closed case  $S_h$ is also concave in the last stage of approaching equilibrium.

\

Regarding $0<\gamma<\frac{2}{3}$, since $\frac{3\gamma-2}{2(\gamma-1)}<1$, an analog analysis shows us that in the last stage of approaching equilibrium $\ddot{S}_{\gamma}>0$ both in $\kappa=0$ and $\kappa\neq0$. %, which means that in this case the thermal equilibrium is not reached.

\

Thus, we need to look at the total second derivative, which has the asymptotic form at the last stage of approaching equilibrium 

\begin{equation}
\ddot{S}=\frac{6\pi\gamma M_{Pl}^2}{a^2\sqrt{H^2+\frac{\kappa}{a^2}}}\left(\frac{4\pi}{\sqrt{H^2+\frac{\kappa}{a^2}}}-\frac{2-3\gamma}{T}\right)\left[-\frac{\gamma\Gamma_0^2}{6}a^2\left(1-\frac{\Gamma_0}{3H} \right)+\kappa \right]
\end{equation}

We observe clearly that the therm in parenthesis is positive as was already justified for equation \eqref{dstot}. We also already know that the therm in claudators is negative. Whereas it is trivial for $\kappa=0,-1$, it is also true for $\kappa=1$ as was justified for equation \eqref{ddshk1}. Therefore, we have proved that for $0<\gamma<2/3$ it also results that $\ddot{S}<0$.

\

To end the case of a constant rate, we take into account quantum corrections, which lead to the following entropy of black hole horizons (see for instance \cite{MP13})
\begin{equation}
S_h=\left[\frac{\mathcal{A}}{4l_{\textit{Pl}}^2}-\frac{1}{2}\ln\left(\frac{\mathcal{A}}{l_{\textit{Pl}}^2} \right)\right]
\end{equation}

Thus, we obtain that:
\begin{equation}
\dot{S}_h=\frac{3\gamma}{2}\left(H-\frac{\Gamma}{3}\right)\left[ \frac{\mathcal A}{2l_{Pl}^2}-1\right]
\end{equation}
which is positive for $H>\frac{\Gamma}{3}$ and $r_h>\frac{l_{Pl}}{\sqrt{2\pi}}$. Taking the second derivative, one gets
\begin{equation}
\ddot{S}_h={9\gamma^2 r_h^2}\left(1-\frac{\Gamma_0}{3H}\right)\left[\frac{\pi}{2l_{Pl}^2}\left(H^2\left(1-\frac{2\Gamma_0}{3H}\right)-\frac{\kappa}{a^2} \right)+ \frac{1}{4r_h^4} \right]+
3\gamma\frac{\kappa}{a^2}\left(\frac{\mathcal A}{4l_{Pl}^2}-\frac{1}{2} \right),
\end{equation}
which near $H=\frac{\Gamma_0}{3}$ becomes 
\begin{equation}
\ddot{S}_h=9\gamma^2\left(1-\frac{\Gamma_0}{3H}\right)\left(-\frac{\pi}{2l_{Pl}^2} +\frac{\Gamma_0^2}{36}\right)+
3\gamma\frac{\kappa}{a^2}\left(\frac{9\pi}{\Gamma_0^2l_{Pl}^2}-\frac{1}{2} \right).
\end{equation}

Then, since $\Gamma_0\ll 1$, in the last stage of approaching equilibrium, for $\kappa=0,-1$ we easily deduce that $S_h$ is concave. For $\kappa=1$, we again need to look at the expression $a^2\left(1-\frac{\Gamma_0}{3H}\right)$. Hence, as in the former cases, we obtain that it is always greater than 1 and, so, in this case it will also be concave.

\subsection[]{Variable creation rate ($\Gamma(H)=-\Gamma_0+3H+\frac{n}{H}$)}

In this case, we proceed analogously as with constant creation rate. The derivative of the entropy of the apparent horizon is:
\begin{equation}
\dot{S}_h=\frac{{\mathcal A}\gamma}{4l_{Pl}^2}\left(\Gamma_0-\frac{n}{H} \right),
\end{equation}
and
the thermodynamic equilibrium $H=\frac{\Gamma}{3}$ takes place at $H=\frac{n}{\Gamma_0}$. And the second derivative of $S_h$ turns out to be
\begin{equation}
\ddot{S}_h= \frac{{\mathcal A}\gamma }{4l_{Pl}^2}\left[\gamma\left(\Gamma_0-\frac{n}{H}\right)\left(\Gamma_0-\frac{3n}{2H}-\frac{n}{2H^3}\frac{\kappa}{a^2} \right) +\frac{n}{H^2}\frac{\kappa}{a^2}\right].
\end{equation}

%\begin{equation}
%\ddot{S}_h=   \left[\gamma\left(\Gamma_0-\frac{n}{H}\right) +{\kappa}\right].
%\end{equation}

Near $H=\frac{n}{\Gamma_0}$, this second derivative becomes
\begin{equation}
\ddot{S}_h= \frac{{\mathcal A}\gamma\Gamma_0^2}{4na^2l_{Pl}^2} \left[-\frac{\gamma n a^2}{2\Gamma_0}\left(\Gamma_0-\frac{n}{H}\right) +{\kappa}\right].
\end{equation}

Consequently, once again, for $\kappa=0,-1$, we have that $\ddot{S}_h<0$ in the last stage of approaching equilibrium. And for $\kappa=1$, we have to analyze what happens with $-\frac{\gamma na^2}{2\Gamma_0}\left(\Gamma_0-\frac{n}{H}\right)+1$ near $H=\frac{n}{\Gamma_0}$.

\

Defining $g=a^2\left(\Gamma_0-\frac{n}{H} \right)$, one can check that, near $H=\frac{n}{\Gamma_0}$, $g$ satisfies the equation
\begin{equation}
\dot{g}=g\left(\frac{2n}{\Gamma_0}-\frac{\gamma\Gamma_0}{2}\right)+\frac{\Gamma_0^2}{n}.
\end{equation}

Therefore, $\lim\limits_{t\to\infty}g(t)=\infty$ if $4n-\gamma\Gamma_0^2\geq0$ and $\lim\limits_{t\to\infty}g(t)=\frac{2\Gamma_0^3}{n(\gamma\Gamma_0^2-4n))}$ if $4n-\gamma\Gamma_0^2<0$. Hence, when $4n-\gamma\Gamma_0^2\geq0$ it follows trivially that, for the closed case, ${S}_h$ is concave in the last stage of approaching the equilibrium. And when $4n-\gamma\Gamma_0^2<0$, we also obtain concavity,
 since near $H=\frac{n}{\Gamma_0}$ one has 
 $-\frac{\gamma n }{2\Gamma_0}g+1<0$ . 
 
 % $\frac{n\gamma}{2\Gamma_0}a^2\left(\Gamma_0-\frac{n}{H}\right)=\frac{\gamma\Gamma_0^2}{\gamma\Gamma_0^2-4n}>1$.

\

If we now proceed to analyze the entropy $S_{\gamma}$ of the matter, we obtain the same expressions as in the constant creation rate for $\dot{S}_{\gamma}$ and $T$:

\begin{equation}
T\dot{S}_{\gamma}={2\gamma\pi M_{Pl}^2r_h}\left(\Gamma_0-\frac{n}{H}\right)(3\gamma-2) \qquad \mbox{and} \qquad  
T=T_0\left(\frac{r_{h,0}}{r_h}\right)^{\frac{2(\gamma-1)}{\gamma}}.
\end{equation}

Thus, as in the previous case, we have that for $H>\frac{n}{\Gamma_0}$, $\dot{S}_{\gamma}<0$ for $\gamma>\frac{2}{3}$. Analogously as in the constant creation rate case, we need to compute the derivative of the total entropy for the case $\gamma<\frac{2}{3}$:

\begin{equation}
\dot{S}=\frac{2\pi\gamma\left(\Gamma_0-\frac{n}{H}\right)M_{Pl}^2}{\sqrt{H^2+\frac{\kappa}{a^2}}}\left[\frac{4\pi}{\sqrt{H^2+\frac{\kappa}{a^2}}}-\frac{2-3\gamma}{T}\right].
\end{equation}

With an analogous argument to the previous case we obtain that $\dot{S}>0$ for the case $0<\gamma<\frac{2}{3}$.

\

Taking the second derivative, we obtain
\begin{equation}
\ddot{S}_{\gamma}=\frac{\gamma\pi(3\gamma-2)^2 M_{Pl}^2r_h}{T}\left\{\left(\Gamma_0-\frac{n}{H} \right) \left(\Gamma_0-\frac{2(2\gamma-1)n}{(3\gamma-2)H}-\frac{\kappa n\gamma}{a^2H^3(3\gamma-2)} \right)+\frac{2\kappa n}{H^2a^2(3\gamma-2))} \right\}.
\end{equation}

Since $\frac{2(2\gamma-1)}{3\gamma-2}>\frac{4}{3}>1$ for $\gamma>\frac{2}{3}$,  $S_{\gamma}$ is concave for $\kappa=0,-1$ in the last stage of approaching the equilibrium $H=\frac{n}{\Gamma_0}$. For $\kappa=1$, we need to look once again  at $\frac{n\gamma a^2}{2\Gamma_0}\left(\Gamma_0-\frac{n}{H}\right)$ near $H=\frac{n}{\Gamma_0}$, which is greater than $1$ as we already showed for $\ddot{S}_h$. Hence, $S_{\gamma}$ is also concave for $\kappa=1$ in the last stage of approaching equilibrium for $\gamma>2/3$.

\

For $0<\gamma<\frac{2}{3}$, given that $\frac{2(2\gamma-1)}{3\gamma-2}<1$, we easily verify that $\ddot{S}_{\gamma}>0$ for both $\kappa=0$ and $\kappa\neq 0$. As a consequence, we need to look again at the second derivative of the total entropy near $H=\frac{n}{\Gamma_0}$:

\begin{equation}
\ddot{S}=\frac{2\pi\gamma M_{Pl}^2\Gamma_0^2}{a^2n\sqrt{H^2+\frac{\kappa}{a^2}}}\left(\frac{4\pi}{\sqrt{H^2+\frac{\kappa}{a^2}}}-\frac{2-3\gamma}{T} \right) \left[-\frac{\gamma n a^2}{2\Gamma_0}\left(\Gamma_0-\frac{n}{H}\right) +{\kappa}\right].
\end{equation}

With the same arguments as in the previous cases we justify that the therm in parenthesis is positive and the one in claudators is negative and, thus, $\ddot{S}<0$ for $0<\gamma<\frac{2}{3}$.

\

When applying quantum corrections, we obtain the same expression as in constant creation rate case for $\dot{S}_{\gamma}$:
\begin{equation}
\dot{S}_h=\frac{\gamma}{2}\left(\Gamma_0-\frac{n}{H}\right)\left[\frac{\mathcal A}{2l_{Pl}^2}-1\right],
\end{equation}
which is positive for $H>\frac{n}{\Gamma_0}$ and $r_h>\frac{l_{Pl}}{\sqrt{2\pi}}$. And the second derivative is:
\begin{equation}
\ddot{S}_h={\gamma^2r_h^2}\left(\Gamma_0-\frac{n}{H}\right)\left[\frac{\pi}{l_{Pl}^2}\left(\Gamma_0-\frac{3n}{2H}-\frac{n}{2H^3}\frac{\kappa}{a^2} \right) +\frac{n}{4H^3r_h^4}\right]+
\frac{n\gamma}{H^2}\frac{\kappa}{a^2}\left(\frac{\mathcal{A}}{4l_{Pl}^2}-\frac{1}{2} \right),
\end{equation}
which
near $H=\frac{n}{\Gamma_0}$ becomes
\begin{equation}
\ddot{S}_h=\frac{\Gamma_0^3\gamma^2}{n^2}\left(\Gamma_0-\frac{n}{H}\right)\left[-\frac{\pi}{2l_{Pl}^2}+\frac{n^2}{4\Gamma_0^2}\right]+
\frac{\Gamma_0^2\gamma}{n}\frac{\kappa}{a^2}\left(\frac{\pi\Gamma_0^2}{n^2l_{Pl}^2}-\frac{1}{2} \right).
\end{equation}

Then, since $\frac{n}{\Gamma_0}\ll 1$, in the last stage of approaching equilibrium, for  $\kappa=0,-1$, $S_h$ is concave. For $\kappa=1$, once again we need to look at the expression $\frac{n\gamma}{2\Gamma_0}a^2\left(\Gamma_0-\frac{n}{H}\right)$. Hence, as in the former cases we obtain that it is always greater than 1 and, so, it will also be concave.

%\small{\nocite{*} \bibliography{refs}}

\

\section{Conclusions}
We have studied two models in cosmologies with adiabatic particle production. The first one, introduced by the time in \cite{Prigogine88}, is the simplest one because the rate of particle production is constant. The other one contains a variable creation rate and is important because it leads to solutions that unify the early inflation with the current cosmic acceleration. First of all, we have performed the dynamical analysis for both models in the FLRW spacetime (spatially flat and curved), showing that, in the expanding phase, all the solutions of both models depict a universe accelerating at late time. The difference appears at early times, in which for a constant creation rate all the solutions in the expanding phase present a singularity at finite cosmic time (the scalar curvature diverges). However, the other model contains non-singular solutions in finite cosmic time that unify the early and late time acceleration. Finally, we have performed the thermodynamical analysis, showing that %when the matter component has an Equation of State parameter, namely $\omega\equiv\gamma-1$, greater than $-\frac{1}{3}$,
all the solutions in the expanding phase with an effective Equation of State parameter $\omega_{eff}$ greater that $-1$ tend asymptotically to the thermal equilibrium.

\

\section*{Acknowledgments}
This investigation has been supported in part by MINECO (Spain), project MTM2014-52402-C3-1-P.

\end{document}